# Soft GPGPUs for Embedded FPGAs: An Architectural Evaluation

Kevin Andryc, Tedy Thomas, and Russell Tessier
*Department of Electrical and Computer Engineering*
*University of Massachusetts*
*Amherst, MA 01003*

*Abstract*—We present a customizable soft architecture which allows for the execution of GPGPU code on an FPGA without the need to recompile the design. Issues related to scaling the overlay architecture to multiple GPGPU multiprocessors are considered along with application-class architectural optimizations. The overlay architecture is optimized for FPGA implementation to support efficient use of embedded block memories and DSP blocks. This architecture supports direct CUDA compilation of integer computations to a binary which is executable on the FPGA-based GPGPU. The benefits of our architecture are evaluated for a collection of five standard CUDA benchmarks which are compiled using standard GPGPU compilation tools. Speedups of $44\times$, on average, versus a MicroBlaze microprocessor are achieved. We show dynamic energy savings versus a soft-core processor of 80% on average. Application-customized versions of the soft GPGPU can be used to further reduce dynamic energy consumption by an average of 14%.

## 1. Introduction

FPGAs are used in a wide variety of embedded systems, such as automotive applications, appliances, and other consumer products. Most of the processing is performed by low-end embedded microprocessors and FPGAs. In some cases, just an FPGA is used and one or more microprocessors are fashioned from FPGA logic to execute specific code types. The benefits of this approach include the ability of software designers to specify functionality in a familiar high-level language (e.g. C) and the flexibility to modify this functionality for the FPGA device without the need to recompile FPGA logic, a time-consuming process that can range from minutes to days.

This paper focuses on an exploration of soft GPGPU architectures in FPGAs. We describe the architectural customization and scalability of FlexGrip (FLEXible GRaphIcs Processor for general-purpose computing), a fully CUDA binary-compatible integer GPGPU, optimized for FPGA implementation [1]. Specifically, we focus on expanding our architecture to include multiple multiprocessors per GPGPU and optimizing away architectural features which are not needed by classes of applications. In developing the soft GPGPU, a series of FPGA-specific optimizations are used. These optimizations, which include the effective use of block RAMs and DSP blocks, are critical to FlexGrip performance. Specific contributions of our work include: (1) We characterize benchmarks into classes and analyze tradeoffs as we vary the amount of conditional execution hardware, number of processor operands and functions supported by the processors. These characterizations allow for the optimization of area and energy and (2) we consider FPGA performance tradeoffs as the number of processors and multiprocessors in the soft GPGPU are varied.

## 2. Background and Related Work

Our soft GPGPU is part of a larger trend in FPGA usage to eliminate long FPGA compile times and difficult hardware design cycles for many designers. Instead of application-specific custom hardware, an architectural overlay [2] is implemented in FPGA hardware. Although these architectures exhibit lower performance and higher energy consumption than their full custom counterparts, they can be swapped into the FPGA on-demand, providing flexibility. For example, over the past ten years, the implementation of soft vector processors on FPGAs has matured significantly [3] [4]. These architectures typically support a customizable number of operations performed in parallel, an optimized memory interface, and a compiler. FPGA usage also allows for the customization of the soft vector processor instruction set and data bit widths [4]. A recent project [3] exploited the pipeline parallelism found in FPGAs to create custom modules that can be integrated into the soft vector processor datapath.

Several FPGA-targeted projects considered the mapping of GPGPU applications represented in OpenCL to multi-threaded FPGA implementations. Labrecque and Steffan [5] described the multithreading of a single processor core. Hazard logic is removed from the processor and hazards are avoided by switching between up to seven different threads. Another work [6] considered an extension of this idea to include multiple cores of these simple multi-threaded processors operating in parallel. Kingyens and Steffan [7] described a GPU-like architecture that has some similarities to our architecture. Their GPU-like architecture includes multithreading across 32 "batches", small cores which contain ALUs. In general, these architectures do not scale to multiple independently-controlled multiprocessors or offer the opportunity for customization based on the specific instructions required by multithreaded applications.

Many recent projects, including commercial offerings, have examined *synthesizing* designs specified in CUDA and OpenCL to application-specific circuits implemented in FPGAs. The MARC architecture [8], a multi-core with custom datapaths, was optimized on a per-application basis to achieve competitive performance versus full-custom FPGA implementation. The FCUDA project [9] developed a tool which converts CUDA programs to a synthesizable version of C. A high-level synthesis tool and FPGA compiler then converts this code to hardware circuits. Shagrithaya *et al.* [10] developed an OpenCL compiler with a library that supports the OpenCL host API. Finally, Altera has developed an OpenCL compiler [11] which converts OpenCL programs to a series of custom parallel compute cores. These efforts achieve excellent performance, energy, and area results at the cost of long compile times.

This paper builds on our basic FlexGrip single multi-processor GPGPU overlay [1]. Previously, we introduced FlexGrip and provided basic details of its architecture, compilation environment, and scalability. In this follow-on work, we describe the effects of architectural optimizations including reducing the numbers of functional units, conditional execution hardware, and memory interfaces on energy consumption. The effects of using multiple multiprocessors to perform computation are also explored. Results for each of these experiments versus a baseline FlexGrip architecture are presented to quantify the results of the optimizations.

## 3. FlexGrip System Overview

### 3.1. FlexGrip System Architecture

We provide a brief discussion of the FlexGrip architecture in this section to motivate our architectural optimizations. A full description of the FlexGrip architecture and host interface can be found in [1].

A *thread block* represents a collection of operations which can be performed in parallel. The kernel instructions and parameters (thread blocks, grid dimensions, etc.), data, control and status are communicated to FlexGrip through a driver via the AXI bus. After initialization, control flow is passed to the GPGPU to execute the CUDA kernel. FlexGrip follows a single instruction, multiple-thread (SIMT) model in which an instruction is fetched and mapped onto multiple *scalar processors* (SPs) simultaneously. The block scheduler is responsible for scheduling thread blocks in a round-robin fashion. The number of thread blocks scheduled at the same time is determined by the number of scalar processors in a *streaming multiprocessor* (SM) and the total number of SMs. After scheduling the thread blocks, the block scheduler signals the warp unit to initiate scheduling the *warps*, where a warp is a smaller set of simultaneous operations, some of which may be performed conditionally. The maximum number of thread blocks that can be scheduled to a SM is restricted by the available shared memory and SM registers.

The GPGPU controller acts as the interface between the block scheduler and the SM. It initializes registers in the vector register file with respective thread IDs.

### 3.2. FlexGrip Streaming Multiprocessor

For this custom FPGA implementation we have developed a five-stage pipelined SM architecture, shown in Fig. 1. The SM includes Fetch, Decode, Read, Execute and Write stages. The *warp unit* at the front of the pipeline coordinates the execution of instructions through the pipeline. Once the block scheduler assigns thread blocks to a specific SM, the warp unit assigns threads to specific scalar processors (SP). This unit schedules warps in a round-robin fashion. Each warp includes a program counter (PC), a *thread mask*, and state. Each warp maintains its own PC and can follow its own conditional path. The mask is used to prevent thread execution within a warp for threads which do not meet specific conditions. Within a warp, threads are arranged in rows depending on the number of scalar processors (SP) instantiated within an SM. For example, for an 8-SP configuration, a warp with 32 threads would be arranged in four rows with each row containing 8 threads. Similarly, for a 16-SP configuration, a warp would be arranged in two rows with 16 threads each. The maximum parallelism is achieved with 32 SPs and one row.

The Fetch stage is the initial stage of the execution pipeline and is responsible for fetching four or eight-byte CUDA binary instructions from system memory. After fetching the instruction, the PC value is incremented (by 4/8 bytes) to point to the next instruction. The Decode stage decodes the binary instruction to generate several output tokens such as the operation code, predicate data, source and destination operands. In the Read stage, source operands are read from the vector register file or shared/global memory blocks depending on the decoded inputs. The vector register file is partitioned, with each thread assigned a set of general-purpose registers. The address register file stores memory addresses for load and store instructions. The Execute stage consists of multiple scalar processors and a single control flow unit. This unit operates on control flow instructions such as branch and synchronization instructions which are described in more detail in the next section. Each thread is mapped to one scalar processor, enabling parallel execution of threads. The Write stage stores intermediate data in the vector register file, memory addresses in the address register file, and predicate flags in the predicate register file. Final results are stored in the global memory.

## 4. Architectural Optimizations

### 4.1. Conditional Branch Optimization

A key contribution of the soft FlexGrip GPGPU is its ability to support thread-level branching in hardware. These resources provide an opportunity for architectural optimization for specific classes of applications which may exhibit less control-intensive behavior. The execution of threads in a warp diverges if the results of a conditional operation are different for different threads. In case of divergence, execution for some threads proceeds along one path (e.g., not-taken) while other threads are idle. When instructions for the not-taken path complete, the execution switches to the alternative execution path (taken path) for the remaining threads while the first set of threads are idle. When both execution paths are finished, a reconvergence point in the code is reached. At this point, execution is resynchronized across all threads and the same set of instruction operations is unconditionally performed by all threads once again.

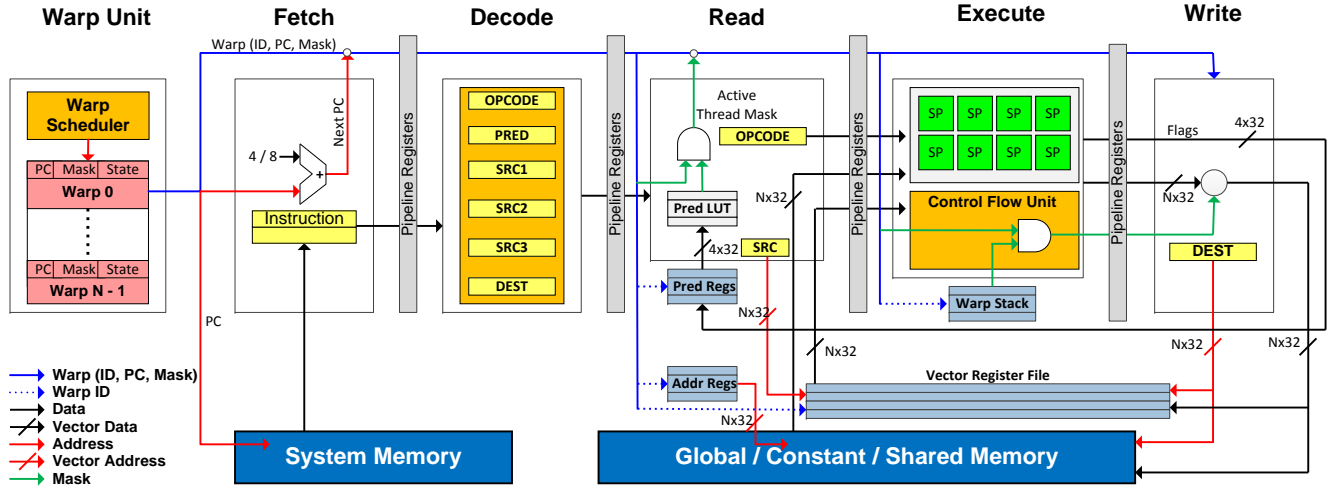

Figure 1. Block diagram depicting the details of the FlexGrip Streaming Multiprocessor

To handle conditional execution, each of the eight warps per SM has its own *warp stack* that includes an instruction address (32 bits), type identifier (2 bits), and an active-thread mask (32 bits) in each stack entry [12] (Fig. 2). The instruction address of the taken branch and the active-thread mask prior to evaluation of the conditional operation are stored on a warp stack for safekeeping. The stored mask contains one bit for each thread in the warp and the type identifier indicates if the instruction address is a reconvergence point or the start address of taken branch instructions. When the taken path of the branch is reached, the stack is popped and the active-thread mask for the warp is inverted to allow for execution of this second path. When the reconvergence point is reached, the original active-thread mask is retrieved by popping the stack.

A complete view of the hardware architecture used to control conditional execution in FlexGrip is shown in Fig. 2. The execution of a conditional (predicate) instruction results in the generation of a four-bit predicate for each instruction (sign, zero, carry, and overflow). This four-bit instruction result for each thread is assigned to a predicate register. Each thread has 4 four-bit predicate registers ($p0$ through $p3$) assigned to it. For each thread, the value in the selected predicate register and the condition for the instruction executed for the branch (e.g. $<, >, =$) are used as in index into a lookup table to generate an instruction mask. One mask bit is generated for each thread. This mask is combined with a thread mask (e.g. thread not *Finished* or *Waiting*) to generate the active-thread mask for the warp. Warp stack pushing and popping of this information is controlled by the control flow unit state machine.

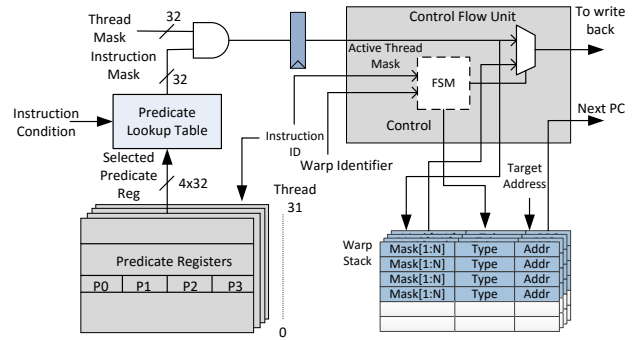

Figure 2. FlexGrip conditional branch and warp stack architecture. There is one stack and one set of predicate registers for each of the eight warps

In the GPGPU control architecture, nested conditionals are possible, requiring a deep stack to hold nested address and mask information. In the worst case only one of 32 threads may execute at a specific time, requiring support for conditional nesting up to 32 entries deep. However, for many applications, a much smaller stack depth is required. This depth can be determined by examining the amount of control nesting in the program or by profiling the application with representative data sets. In our optimizations, we consider the application warp stack depth as an optimization parameter.

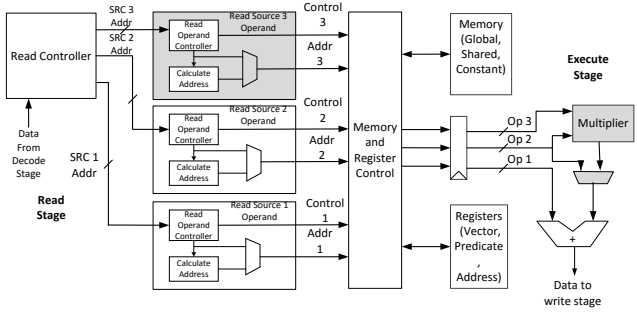

Figure 3. FlexGrip read stage and execute unit

### 4.2. Source Operand Optimization

Fig. 3 depicts the detailed view of the read stage which consists of a read controller, parallel read source operand units, and interface controllers to memory subsystems and registers. The arithmetic portion of the execute stage is shown on the right of the figure. The read controller takes in data from the decode stage, performs pre-processing depending on the operation, and then directs the data to each of the read operand units. These units are functionally identical, allowing for read operations to be performed in parallel. However, they can perform different functions depending on the instruction passed to them at run time. For example, one of the modules may perform a read operation from global memory, while the others perform a read operation from the register file.

The modular independence of the read hardware allows for the removal of one of the read operand modules and the multipler if they are not needed by an application. For example, if an application that does not perform multiply or multiply-accumulate operations, a version of the GPGPU which does not include these features could be used. This hardware is represented by the shaded blocks in Fig. 3. The area and energy benefits of removing this hardware for selected applications is explored in Section 5.

### 4.3. Multiple Streaming Multiprocessors

A notable feature of our architecture is its support for multiple SMs. A thread block of up to 256 threads can be assigned to any available SM by the block scheduler. The number of thread blocks is specified by the programmer and passed to the FlexGrip architecture by the MicroBlaze driver at run-time. The allocation of SM shared memory and the number of registers required per block are also determined during scheduling. The values are determined during compilation and stored in GPGPU configuration registers. After assignment by the block scheduler, the warp unit in the SM uses the parameters to generate and schedule warps.

At the start of kernel execution, the maximum number of thread blocks that can be scheduled is calculated. This value is limited by the number of allocated warps per SM, the number of registers per SM, and the size of the shared memory per SM. Table 1 lists the physical limits of the FlexGrip GPGPU. Control signals from the SM notify the block scheduler when all threads blocks have completed and scheduling of subsequent blocks can begin. Once all thread blocks have successfully executed, the block scheduler signals the GPGPU which will notify the driver that execution has completed.

TABLE 1. FLEXGRIP PHYSICAL LIMITS

| Parameters | Constraint |
|---|---|
| Threads Per Warp | 32 |
| Warps Per SM | 24 |
| Threads Per SM | 768 |
| Thread Blocks Per SM | 8 |
| Total Number of 32-bit Registers per SM | 8,192 |
| Shared Memory Per SM (bytes) | 16,384 |

## 5. Experimental Results

The soft GPGPU supports the NVIDIA G80 instruction set with compute capability 1.0. We tested 27 integer CUDA instructions as a part of this research. All instructions needed by our benchmark circuits are supported. All circuitry described in this section has been implemented in a Virtex-6 FPGA and has been shown to operate correctly. Synthesis was performed using the Xilinx ISE 14.2 toolkit and Modelsim SE 10.1 was used for simulation and verification.

We have evaluated five CUDA applications, *bitonic sort, autocorrelation, matrix multiplication, parallel reduction* and *transpose* from the University of Wisconsin ERCBench suite [13] and the NVIDIA Programmer's Guide [14], using FlexGrip. Additional information regarding the compilation environment and benchmarks is available in [1].

### 5.1. Comparison versus the Microblaze Soft-Core Processor

The FlexGrip soft GPGPU design was implemented on a Xilinx ML605 development board which utilizes a Virtex-6 VLX240T device. The device area and design operating frequency for designs with a varying number of scalar processors and streaming multiprocessors are annotated in Table 2. All designs were evaluated at 100 MHz. We performed experiments and compared performance and energy results against a Xilinx MicroBlaze soft-core processor with 3,252 LUTs running at 100 MHz using C versions of the same benchmarks. For the purposes of this paper, a design with a single SM and 8 scalar processors was implemented and benchmarked on the ML605 board, while 1 SM, 16- and 32-SP designs were simulated. We also extended the design to compare a single SM versus two SMs, each with 8, 16, and 32-SP via simulation. The same baseline FlexGrip design with no architectural optimizations implemented in hardware could successfully run all five benchmarks using the same FPGA bitstream. The CUDA compile times for all benchmarks were less than one second.

TABLE 2. AREA COMPARISON OF BASELINE FLEXGRIP IMPLEMENTATIONS

| Parameters | LUTs | FFs | BRAM | DSP48E |
|---|---|---|---|---|
| 1 SM - 8 SP | 60,375 | 103,776 | 124 | 156 |
| 1 SM - 16 SP | 113,504 | 149,297 | 132 | 300 |
| 1 SM - 32 SP | 231,436 | 240,230 | 156 | 588 |
| 2 SM - 8 SP | 135,392 | 196,063 | 238 | 306 |
| 2 SM - 16 SP | 232,064 | 287,042 | 262 | 594 |
| 2 SM - 32 SP | 413,094 | 468,959 | 310 | 1170 |

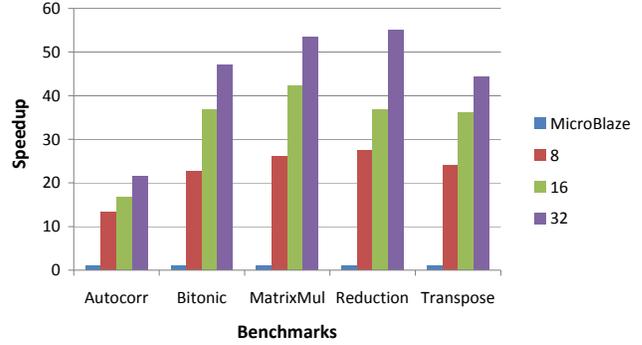

Figure 5. Speedup vs. MicroBlaze for variable scalar processors and input data size 256 for 2 SM

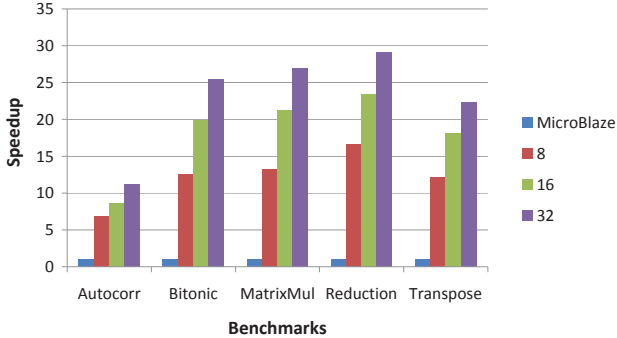

Figure 4. Speedup vs. MicroBlaze for variable scalar processors and input data size 256 for 1 SM

**5.1.1. Architecture Scalability.** We ran experiments by varying the number of scalar processors within a single SM and across 2 SMs which effectively varies the number of threads that can be executed in parallel. Benchmarks *autocorr*, *bitonic*, and *reduction* used input data sets of 32, 64, 128, and 256 values. Benchmarks *matrix multiplication* and *transpose* used input data sets of 32×32, 64×64, 128×128, and 256×256 for experimentation. Fig. 4 shows application speedups versus a MicroBlaze for a varying number of SPs per SM for an input size of 256 (or 256×256) on a single SM. Application speedups range from 7× to 29× with an average speedup close to 12× for 8 SPs, 18× for 16 SPs, and 22× for 32 SPs.

To explore additional speedup, experiments were also performed with 2 SM with 8-, 16-, and 32-SP each. The block scheduler logic equally and automatically distributed thread blocks to the 2 SMs, thus reducing the workload of each SM to roughly half of the 1 SM cases. All benchmarks exhibited additional speedups versus the 1 SM case for the same number of SPs per SM. As shown Fig. 5, the peak speedups for the 2 SM, 32-SP implementations of the benchmarks offer over a 40× speedup for four out of the five benchmarks. Table 3 shows the scalability of our architecture. Speedups for 2 SM versus 1 SM versions of the same benchmark ranged from 1.77 (*Reduction*) to 1.98 (*Transpose* and *Matrix Multiply*). The block scheduler was able to distribute thread blocks more evenly between the two SMs for the latter two applications due to a smaller number of conditional statements in the applications versus the other three applications.

TABLE 3. SPEEDUP OF 2 SM VERSUS 1 SM FOR INPUT DATA SIZE 256

|  | 8 SP | 16 SP | 32 SP |
|---|---|---|---|
| Autocorr | 1.94 | 1.94 | 1.94 |
| Bitonic | 1.82 | 1.83 | 1.85 |
| MatrixMul | 1.98 | 1.98 | 1.98 |
| Reduction | 1.78 | 1.77 | 1.77 |
| Transpose | 1.98 | 1.98 | 1.98 |

**5.1.2. Energy Efficiency.** We used Xilinx's XPower power estimator tool to determine static and dynamic power for the designs (Table 4). Since static power is largely a function of the device size, we evaluate the dynamic energy consumption of the implementations. This value is determined by multiplying dynamic power by application execution time. In Table 5, it is shown that the baseline FlexGrip dramatically reduces dynamic energy consumption versus the MicroBlaze, primarily due to reduced execution time. FlexGrip also uses the same instruction for many scalar processors, limiting instruction memory accesses. For a 1 SM, 8 SP configuration, the average dynamic energy reduction is about 80%, on average.

## 5.2. Architectural Customization

To evaluate the possible benefits of removing unneeded features from FlexGrip, we ran several experiments to determine the minimum required architectural configuration for area and energy optimization for each applications. As described in Section 4, the specific optimizations include reducing the size of the warp stack (and associated control logic) and removing the multiplier and the third-operand read circuitry from the read stage of the SM pipeline. Table 6 lists the optimizations performed for each of the benchmarks. By performing an instruction analysis, we can determine the minimal set of functions needed to support each benchmark. Of the five benchmarks, we were able to remove the multiplier/third operand for *bitonic*, since the benchmark does not require multiply operations. Effectively, any benchmark which performs multiplies could use this FlexGrip version and obtain the 23% dynamic energy reduction versus FlexGrip with a reduced warp stack and 38% dynamic energy reduction versus baseline FlexGrip. We note that only the multiply-add (MAD) instruction requires three operands, therefore by eliminating the multiply unit the need for support of a third operand is removed. A total of 12 DSP blocks are still used for address calculation in the FlexGrip control circuitry.

TABLE 5. MICROBLAZE VS. FLEXGRIP ENERGY CONSUMPTION: 256 DATA SIZE

|  | MicroBlaze | | 8 SP | | | 16 SP | | | 32 SP | | |
| --- | --- | --- | --- | --- | --- | --- | --- | --- | --- | --- | --- |
|  | Exec. Time (ms) | Dyn. Ene. (mJ) | Exec. Time (ms) | Dyn. Ene. (mJ) | Ene. Red. | Exec. Time (ms) | Dyn. Ene. (mJ) | Ene. Red. | Exec. Time (ms) | Dyn. Ene. (mJ) | Ene. Red. |
| Autocorr | 277.0 | 102.49 | 40.28 | 33.84 | 67% | 32.20 | 34.78 | 66% | 24.89 | 34.60 | 66% |
| Bitonic | 118.0 | 43.66 | 9.39 | 7.88 | 82% | 5.95 | 6.43 | 85% | 4.64 | 6.44 | 85% |
| MatrixMul | 186041.0 | 68835.17 | 14098.02 | 11842.34 | 82% | 8735.90 | 9434.77 | 86% | 6904.07 | 9596.66 | 86% |
| Reduction | 11.0 | 4.07 | 0.66 | 0.55 | 86% | 0.47 | 0.51 | 87% | 0.38 | 0.53 | 87% |
| Transpose | 705.0 | 260.85 | 57.79 | 48.54 | 81% | 38.74 | 41.84 | 84% | 31.48 | 43.75 | 83% |

TABLE 4. FPGA POWER ESTIMATES (W) AT 100 MHZ

|  | Dynamic | Static | Total |
| --- | --- | --- | --- |
| 1 SM, 8 SP | 0.84 | 3.45 | 4.29 |
| 1 SM, 16 SP | 1.08 | 3.46 | 4.54 |
| 1 SM, 32 SP | 1.39 | 3.46 | 4.85 |
| MicroBlaze | 0.37 | 3.45 | 3.82 |

TABLE 6. RESULTS OF FLEXGRIP OPTIMIZATIONS FOR AN 1 SM, 8 SP SYSTEM

|  | Num. of Oper. | Warp Depth | Slice LUTs | Flip Flops | Block RAM | DSP | % Area Red. | % Dyn. Red. |
| --- | --- | --- | --- | --- | --- | --- | --- | --- |
| Baseline | 3 | 32 | 60,375 | 103,776 | 124 | 156 | - | - |
| Autocorr. | 3 | 16 | 52,121 | 82,017 | 124 | 156 | 14% | 3% |
| Mat. Mult. | 3 | 0 | 42,536 | 60,161 | 124 | 156 | 30% | 9% |
| Reduction | 3 | 0 | 42,536 | 60,161 | 124 | 156 | 30% | 9% |
| Transpose | 3 | 0 | 42,536 | 60,161 | 124 | 156 | 30% | 9% |
| Bitonic | 3 | 2 | 39,189 | 57,301 | 124 | 156 | 35% | 15% |
| Bitonic | 2 | 2 | 22,937 | 27,136 | 120 | 12 | 62% | 38% |

Table 6 indicates that the necessary depth of the warp stack for applications varies. As noted in Section 4, each warp has its own warp stack which is configured as 32 registers of 66-bits each. For short instruction sequences, such as if statements without a corresponding else, the compiler uses condition codes to avoid managing divergence, reducing the need for significant warp stack depth. In cases with longer sequences of conditional code, conditional branches are used. For *matrix multiplication*, *reduction*, and *transpose*, conditional branches are minimized, limiting warp stack usage. By customizing the warp stack, a LUT area reduction of up to 35% and a dynamic energy reduction of up to 15% can be realized.

In an embedded system, one could consider compiling and storing the bitstreams for four separate FlexGrip GPGPUs. The baseline system would include a multiplier and a full 32-depth warp stack. A second system would include a 16-depth warp stack and a third system would have a 2-depth stack. Finally, the fourth system would include a 2-deep warp stack and no multiplier/third operand fetch unit.

## 6. Conclusions

In this paper we explore the possibility of providing a small set of FlexGrip soft GPGPU implementations that could be targeted to classes of applications with different execution characteristics (e.g., reduced conditional operation, no multiplication). We show that architectural optimization can reduce dynamic energy consumption by 14% and LUT area by 33%, on average. Experimental results demonstrate application speedups of up to 55× for a FlexGrip design with two streaming multiprocessors (SMs) versus a MicroBlaze soft processor operating at the same clock frequency for highly parallel benchmarks.